\documentclass[aps,prl,twocolumn,showpacs,floatfix]{revtex4-1}

\usepackage{amsmath}
\usepackage{amssymb}
\usepackage{graphicx}

\usepackage{color}

\begin{document}

\date{\today}

\author{Ganna Berezovska}
\author{Diego Prada-Gracia}
\author{Francesco Rao}
\email{francesco.rao@frias.uni-freiburg.de}

\affiliation{Freiburg Institute for Advanced Studies, School of Soft Matter
Research, Freiburg im Breisgau, Germany.}

\title{Consensus for the Fip35 folding mechanism?}

\begin{abstract}

	Recent advances in computational power and simulation programs finally
	delivered the first examples of reversible folding for small proteins
	with an all-atom description. But having at hand the atomistic details
	of the process did not lead to a straightforward interpretation of the
	mechanism. For the case of the Fip35 WW-domain where multiple long
	trajectories of 100 $\mu$s are available from D. E. Shaw Research,
	different interpretations emerged.  Some of those are in clear
	contradiction with each other while others are in qualitative
	agreement. Here, we present a network-based analysis of the same data
	by looking at the \emph{local fluctuations} of conventional order
	parameters for folding.  We found that folding occurs through two major
	pathways, one almost four times more populated than the other. Each
	pathway involves the formation of an intermediate with one of the two
	hairpins in a native configuration.  The quantitative agreement of our
	results with a state-of-the-art reaction coordinate optimization
	procedure as well as qualitative agreement with other
	Markov-state-models and different simulation schemes provides strong
	evidence for a multiple folding pathways scenario with the presence of
	intermediates.

\end{abstract}

\maketitle

\section{Introduction}

Computer models of protein folding have been around now for almost two decades
\cite{lazaridis1997,dobson1998}. Such models evolved dramatically, going from
simple lattice models \cite{shakhnovich1991} and implicit solvent
implementations \cite{ferrara2000} to fully atomistic calculations
\cite{freddolino2010}. In the quest for microscopic models of protein folding,
several groups focused on small proteins like WW domains which can be very
informative yet much easier to treat.  Recently, an important breakthrough in
this direction was delivered by D.~E.~Shaw Research.  By using in-house
technology with optimized software they delivered first examples of reversible
folding for some mini-proteins  \cite{shaw2010}.  For the case of the Fip35 WW
domain they made  available two trajectories each of which of 100
$\mu$s in length.  These calculations raised a number of controversial
interpretations on the actual folding mechanism. 

For this system, no agreement was found on whether the folding process proceeds
via on-pathway intermediates or downhill folding. The original work applied an
optimization procedure to obtain a reaction coordinate for the folding process
\cite{best2005}, concluding that no relevant barriers are present in the
folding process \cite{shaw2010}. This interpretation was challenged by others.
In particular, Krivov demonstrated the presence of intermediates by calculating
a novel optimized reaction coordinate \cite{krivov2011}.  This view was also
supported, at least at a qualitative level, by Pande and Baker groups which
independently analyzed the same data with Markov-state-models \cite{pande2011,
hua2012}.  

All approaches have found that the folding process is more likely to
proceed via the formation of the first hairpin ($\beta_1$) followed by the second
one ($\beta_2$) but the conceptual disagreement on the presence of
intermediates or a downhill scenario is pretty strong. Unfortunately, these
types of analysis \cite{shaw2010,krivov2011,pande2011,hua2012} are not very intuitive, making an objective evaluation of
the results hard. On one side, optimization procedures like the ones applied by
Shaw and Krivov, tend to hinder the physical meaning of the obtained reaction
coordinates while clustering of high-dimensional spaces strongly suffer
from thermal fluctuations \cite{rao2010} and limited sampling \cite{hua2012}. 

In an effort to bridge the gap between the use of more intuitive coordinates
and the application of Markov-state-models, we recently extended an approach
derived from single-molecule spectroscopy to study molecular simulations
\cite{berezovska2012}.  Aiming at identifying the most robust features of Fip35
folding, we present here an extension of this framework for the analysis of
D. E. Shaw data.  The entire analysis is based on conventional order
parameters time series, like root mean square deviations (RMSD). This overcomes
the problem of working in complex multidimensional spaces as in the case of
coordinate optimizations \cite{shaw2010, krivov2011}, k-means clustering
\cite{pande2011} or contact maps comparisons \cite{hua2012}.  Still, the approach
provides a good assessment of the kinetics thanks to the application of complex
networks analysis, allowing the development of simple Markov-state-models
reproducing the time scales of the original MD trajectory \cite{rao2010,
berezovska2012}. Our results  reinforce  the interpretation that Fip35 folding proceeds via intermediates.

\section{Methods}

\subsection{Markov state models from conventional order parameter analysis}

In this section we are going to explain the tools that were used to analyze the
Fip35 data set. The main idea behind \emph{local fluctuations analysis} is to
build a Markov-state-model using as input the time series of a general order parameter. 
This is done by looking at the fluctuations of the coordinate within a predetermined 
time window. The approach was initially developed for single 
molecule experiments \cite{baba2007,schuetz2010,baba2011}, but in a recent paper we applied
and extended this technique to study conventional order parameter time series
from molecular simulations \cite{berezovska2012}. The main motivation was to
develop a tool to analyze those time series in a more rigorous way, going
beyond straightforward histogram analysis. In fact, the great advantage of this strategy is to
characterize order parameter time series on the base of the kinetics, something
that was definitively impossible by using free-energy projections and histogram
analysis. The approach takes advantage of the work done in complex network
analysis \cite{rao2004,gfeller2007} and Markov-state-models \cite{prinz2011} in
molecular systems but it overcomes some of the problems, e.g. avoiding to work in
highly dimensional spaces. The downside here is that the framework is based on simple
coordinates, therefore  if they miss some relevant aspects of the system there is no way to
recover that type of information.

The steps to be covered are very similar to any other
Markov-state-model: (i) microstate building, (ii) transition network building,
(iii) kinetic lumping. Although already discussed elsewhere  in detail
\cite{rao2004,gfeller2007,berezovska2012}, below we provide the essentials to
better follow the paper. A code to reproduce the presented analysis is freely
distributed at the website {\tt raolab.com}.

\subsubsection{Microstate building} 

Microstates were defined for every trajectory snapshot by looking at the local
fluctuations of the order parameter coordinate within a time window $t_w$
centered in the snapshot itself \cite{schuetz2010,berezovska2012}.  Two time
points belonged to the same microstate if they had comparable distributions of
the order parameter within $t_w$ according to a Kolmogorov-Smirnov test
\cite{KST}. That is, if the condition $ D\leq \zeta \sqrt{2/t_w} $ was
fulfilled, where $D$ is the maximum difference of the two cumulative
distributions and $\zeta$ corresponds to a certain confidence level. Being
$t_w$ and $\zeta$ related, we fixed the latter value to 0.5 and let $t_w$ vary
as done in Ref. \cite{berezovska2012}.  Comparisons were made along the
trajectory using the leader algorithm in a way that every time point was
associated to a microstate \cite{seeber2007, schuetz2010}.  As shown by others
\cite{schuetz2010} and us \cite{berezovska2012}, the methodology is robust for
a reasonable wide range of time windows. For example, for values of the time
window up to 13 ns the mean first passage time to the folded state steadily
increases till a plateau. Then, between 13 and 24 ns the mean first passage time
fluctuates around 4.6 $\mu$s. Finally for larger windows, this value tends to
slightly increase together with larger fluctuations but still in agreement with previous
calculations \cite{krivov2011}.  In the following, we fix the time window to
$18$ ns. In the general case, short time windows are unable to capture
significantly well the coordinate distribution leading to faster kinetics
\cite{berezovska2012} while long ones result in too many fast fluctuations to
neighboring states inside a single distribution.  Given the $\mu$s time scales
of the folding process, this would happen only in cases when windows of
hundreds of ns are selected.

\subsubsection{The configuration-space-network} 

The resulting time series of microstates was mapped onto a
configuration-space-network \cite{rao2004,rao2004, gfeller2007,
gfeller2007uncovering, rao2010jpcb, berezovska2012}.  Microstates represent network
nodes and a link between them exists if they were successively visited along
the molecular trajectory.  Detailed balance was imposed for each link by making
an average of the number of transitions in both directions.  This was only
partially necessary because the original trajectories mostly satisfied detailed
balance already.  

\subsubsection{Kinetic lumping by network clusterization}

Protein conformational states were defined by applying a kinetic lumping
scheme. This was done by running a clusterization procedure on the
configuration-space-network, the Markov-Clustering-Algorithm (MCL)
\cite{enright2002}.  This approach assures that the obtained network clusters
represent meaningful free-energy basins with preserved system kinetics
\cite{gfeller2007,berezovska2012}. Being interested on the characterization of
the folding mechanism, a granularity parameter of $1.2$ was used to focus on the
highest barriers only \cite{gfeller2007,berezovska2012}. 
The obtained states were used to build a Markov-state-model, schematically
representing all relevant slow transitions of the system.

\subsection{First passage time distributions}

The kinetic similarity between the original trajectory and the
Markov-state-model was investigated by comparing the distribution of the
first-passage-times (fpt) \cite{rao2010,berezovska2012} to the folded state.
This is the distribution of times to reach the native state from any
other snapshot of the trajectory \cite{rao2010}. For the Markov-state-model  
the fpt was calculated on  trajectories  generated by running a random walk.  
Arrival times depend on the definition of the target only and not on the 
detailed decomposition of the trajectory. For this study we used the definition 
of the native state as obtained by MCL.

\subsection{Molecular simulation data}

The simulation data was directly obtained from D. E. Shaw Research and published
in Ref.  \cite{shaw2010}  It is an all-atom molecular dynamics simulation  in
explicit water (TIP3P water model) at the protein's in silico melting
temperature ($395 K$). It was calculated using the Anton supercomputer with the
modified Amber ff99SB-ILDN force field \cite{Lindorff-Larsen2010} carried out
in the NVT ensemble using the Nose-Hoover thermostat with a relaxation time of
$1.0$ ps \cite{Lindorff-Larsen2010, shaw2010}. The simulation data consisted of
two trajectories, each of length $100 \,\mu s $.

\section{Results}

\subsection{RMSD analysis of the folding mechanism: identification of putative
intermediate states}

\begin{figure}
\includegraphics[width=80mm,angle=0] {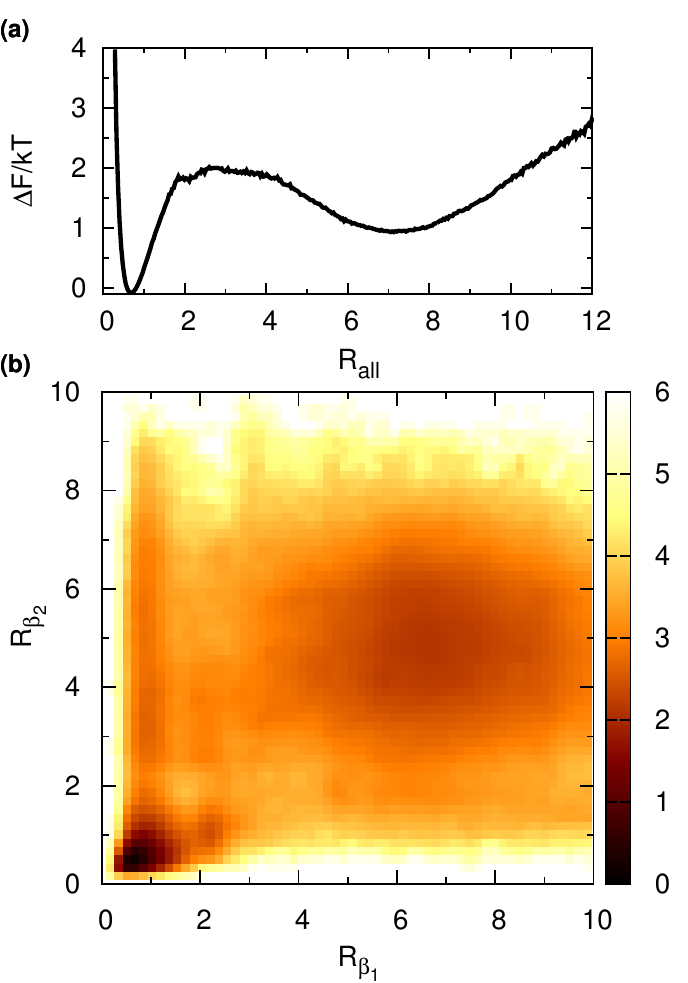}
\caption{Projected free-energy surfaces of Fip35: (a) as a function of the RMSD
coordinate $R_{all}$ (backbone atoms from residues $7-29$); (b) as a function of
the RMSD coordinates $R_{\beta_1}$ (residues $7-23$) and $R_{\beta_2}$ (residues
$18-29$). Free energies are expressed in units of $kT$.}
  \label{fig:projections}
\end{figure}

Conventional order parameters for protein folding include RMSD with respect to
the native state \cite{duan1998}, number of native contacts \cite{ferrara2000}
or radius of gyration \cite{rao2003}.  RMSD is certainly one of the most
obvious choices when it comes to monitor folding to a known structure because
it requires minimal a priori knowledge (i.e. the native structure).  The projected free-energy landscape onto the RMSD from the
native state ($R_{all}$) shows two clear minima, corresponding to the folded and unfolded
states (Fig.~\ref{fig:projections}a). To improve on this simple description we
projected the landscape onto two coordinates instead of one.  Given the
triple stranded topology of Fip35,  we expect that the RMSD coordinates from the
first (residues $7-23$, $R_{\beta_1}$) and second (residues $18-29$, $R_{\beta_2}$) hairpin to provide additional information on the process. Fig.~\ref{fig:projections}b shows a 2D projection onto these two
new coordinates. The folded and unfolded states are clearly visible  at
regions around $(1,1)$ and $(7,5)$ respectively. In agreement with the 1D projection,
this plot provides some further information on the presence of other states
like the darker regions at around $(2,1)$ and $(1,3.5)$. Those regions
might represent intermediate steps to the folding process, a property that  cannot
be validated by Fig.~\ref{fig:projections}b due to the lack of information on the kinetics \cite{rao2004, krivov2004}.

\begin{figure}
\includegraphics[width=80mm] {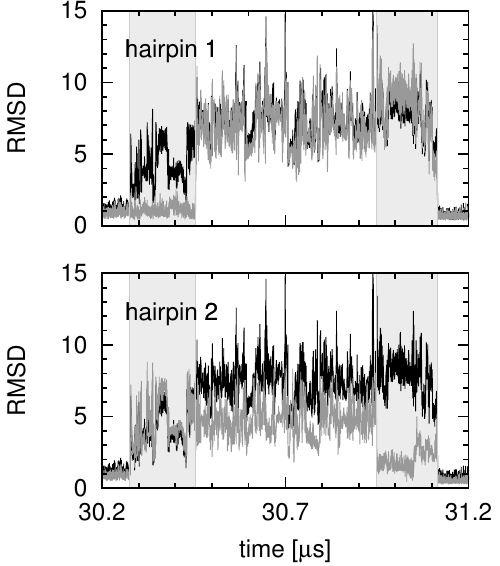}
\caption{A time series stretch of Fip35. The RMSD for the whole structure
  $R_{all}$ is shown in black. $R_{\beta_1}$ and $R_{\beta_2}$ are shown as
  gray lines in the top and bottom panels, respectively.  Gray band regions
  represent portions of the trajectory where one of the two hairpins is native
  while the other one is unfolded.  }
  \label{fig:RMSD}
\end{figure}

The relevance of these regions for the folding process emerged  by
the inspection of some of the folding/unfolding events. One example is shown
in Fig.~\ref{fig:RMSD}. In the two panels, the RMSD with respect to the native
state $R_{all}$, first hairpin ($R_{\beta_1}$, top panel) and second hairpin ($R_{\beta_2}$, bottom panel) are shown as black and gray lines, respectively. This picture shows that at around  $30.3$ $\mu$s 
 an unfolding event is present. Here  $R_{all}$ rapidly
increases (black line) as well as $R_{\beta_2}$ (gray, bottom panel). This is
not the case for $R_{\beta_1}$ (gray, top panel) which stays low for around $200$
ns while the other hairpin is unfolded (the region is highlighted with a gray band). A similar behavior was
observed for the folding event at $31.1$ $\mu$s, with the second hairpin being
native for a time span of roughly $150$ ns before the complete folding event
(right gray band).  In the folding/unfolding process the two hairpins evolve in an uncorrelated manner. 
Consequently, the RMSD coordinates $R_{\beta_1}$ and  $R_{\beta_2}$ provide independent 
information on the folding mechanism, suggesting the presence of on-pathway intermediates. 
It is important to note, while the total RMSD is able to report on the presence of these states (the RMSD in
the gray band regions of Fig.~\ref{fig:RMSD} is lower than the completely
unfolded state), this coordinate does not have the sensitivity to discriminate
between partially folded states with the first hairpin formed from the ones
with the second hairpin formed. For this reason, we do not think that $R_{all}$ represents a good 
coordinate for a local fluctuations analysis.

\subsection{Local fluctuations analysis: kinetics assessment of the
intermediate states}

Given these observations, we chose $R_{\beta_1}$ and $R_{\beta_2}$ coordinates as
probes for a more insightful kinetic analysis of the folding mechanism. This was done by performing 
a joint local fluctuations analysis of these coordinates (see the Methods section for details).
Being this approach developed for a single coordinate, here we
extended the framework to account for multiple order parameters, such as  $R_{\beta_1}$ and $R_{\beta_2}$.
To do so, the local fluctuations of each coordinate were first analyzed separately using the
standard approach. Then, the obtained \emph{states} for each coordinate were merged
into a set of ``combined'' states.  That is, given at a certain time the states corresponding to $R_{\beta_1}$ and $R_{\beta_2}$ being respectively $A$ and $B$, the new combined state is $(A,B)$. This strategy includes the contribution of two coordinates in a simultaneous way.

\begin{figure}
\includegraphics[width=80mm] {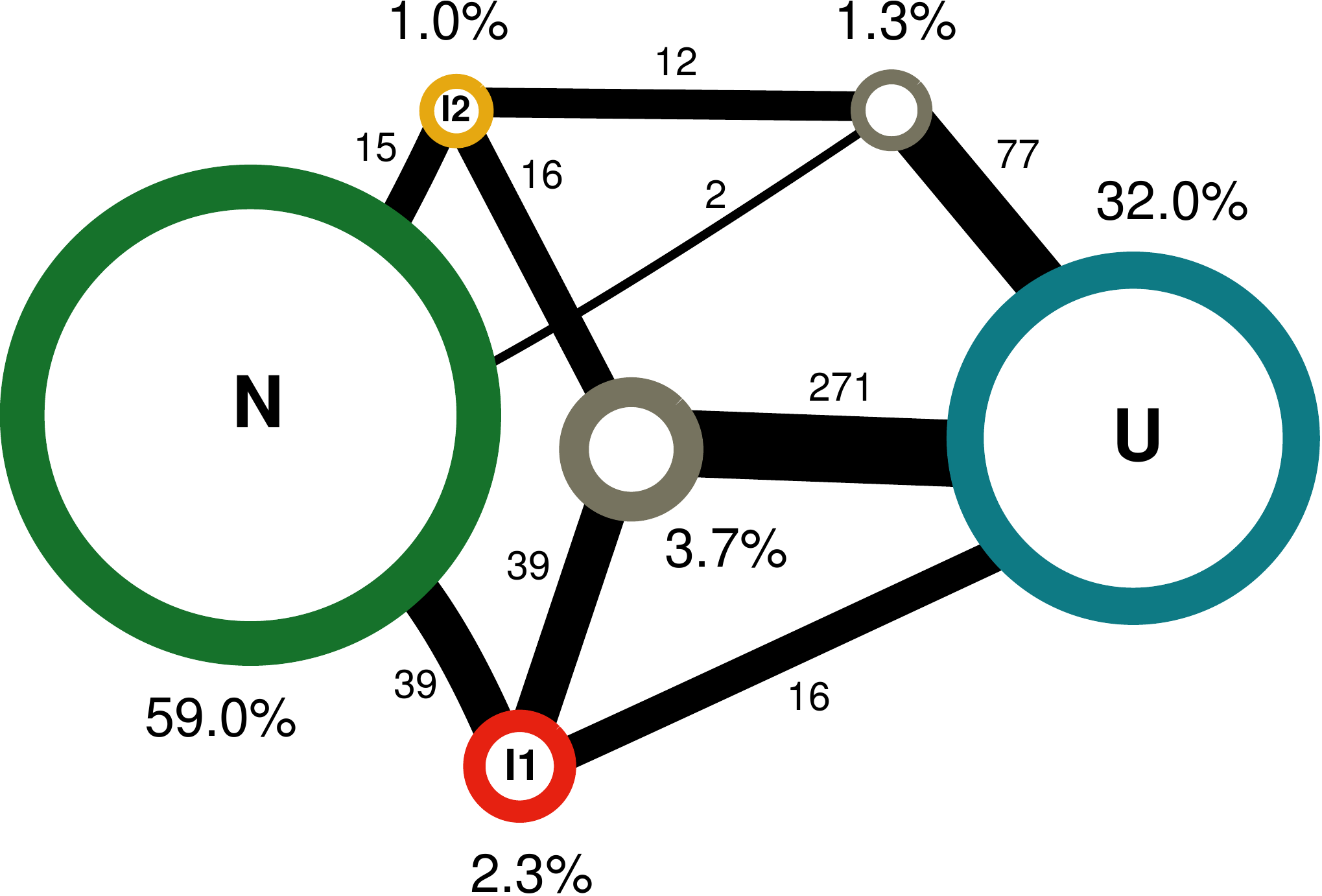}
\caption{Markov-state-model. States were obtained by a local fluctuations
analysis  of $R_{\beta_1}$ and $R_{\beta_2}$ followed by a kinetic lumping via
network clusterization (see Methods for details). Total number of transitions
and state populations are indicated.}
  \label{fig:network}
\end{figure}

Application of this technique resulted in the identification
of six states with a population larger than (or equal to) $1.0 \%$. The
cumulative population of these states is of about $99.3 \%$, indicating that
they well characterize the sampled conformational space. In this representation
the native (N) and unfolded states (U) have a population of $59.0 \%$ and $32.0
\%$, respectively.  The remaining four states have a much smaller population of
few percents. The six states were used to build a reduced Markov-state-model
whose transition network is shown in Fig.~\ref{fig:network}. Interestingly, all
states stay on-pathway from U to N. Specifically, two major folding
intermediates were found just preceding the fully folded state: I1 and I2.
These intermediates correspond to two independent folding routes with different
relative populations. We calculated this explicitly by looking along the
original trajectory which states were preceding the folding state. Of the total
$10$ folding events $7$ and $2$ events followed the I1 and I2 routes,
respectively. A third pathway was followed only once.

\begin{figure}
  \includegraphics[width=80mm, angle=-90] {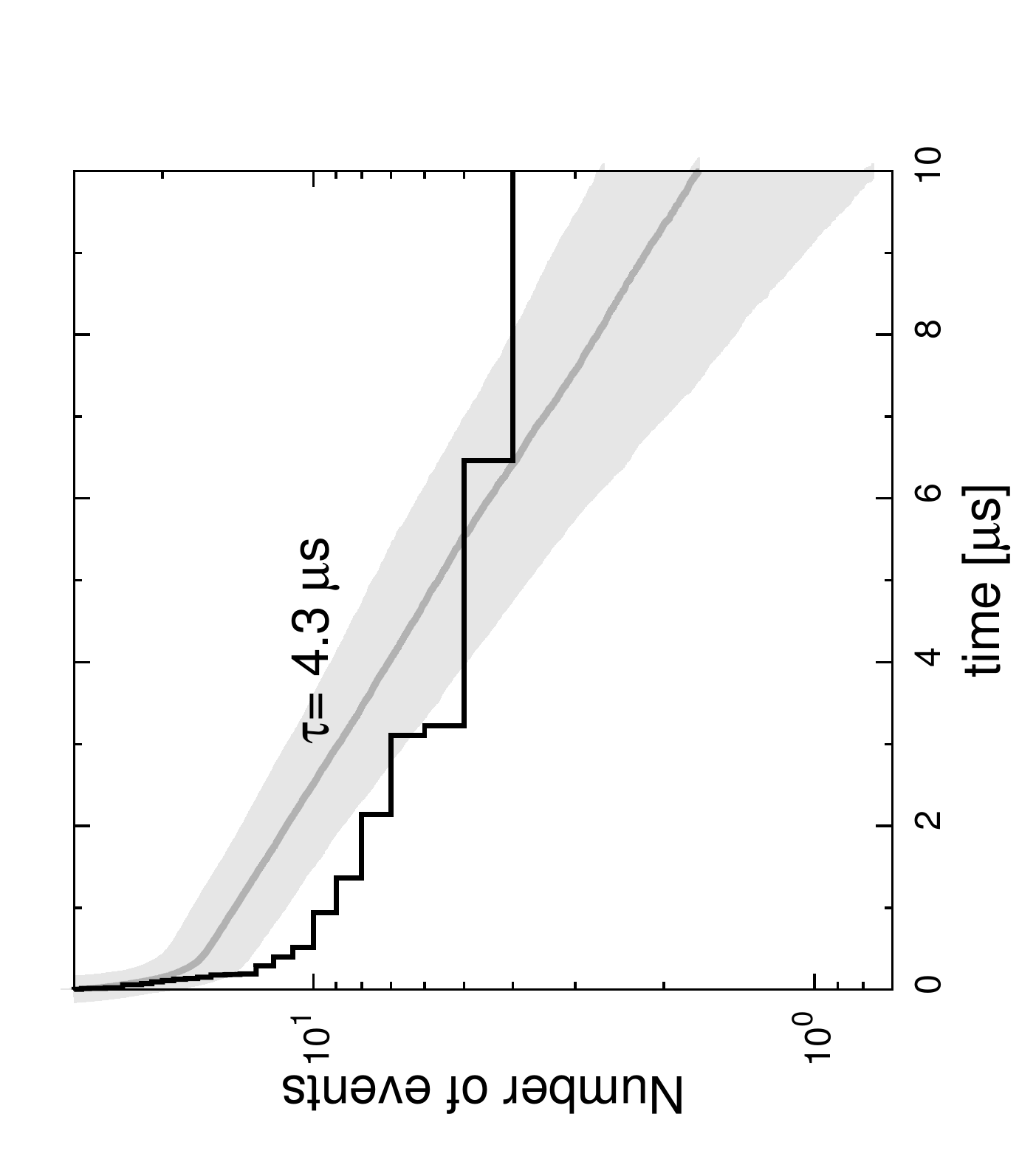}
\caption{First-passage-time distributions to the native state. The black line
corresponds to the first passage time distribution along the original MD
trajectory. The gray line together with the light gray region
respectively show the average first passage time value and its standard deviation obtained by
$10^3$ random walks of length $10^6$ steps on the Markov-state-model of Fig.~\ref{fig:network}. 
An exponential fit of the data showed a relaxation time of
$4.3\, \mu s$ which is in good agreement with the value estimated from
trajectory \cite{krivov2011}. }
  \label{fig:fpt}
\end{figure}

To check that the reduced Markov model of Fig.~\ref{fig:network} was able to
correctly reproduce the original dynamics of the MD trajectory, a first passage
time analysis to the native state was computed.  In Fig.~\ref{fig:fpt} the
distributions of the first passage times corresponding to  the original trajectory and the
six-states Markov model are shown as black and gray lines, respectively.
Interestingly, the two distributions present a very similar decay in the long
times regime, corresponding to a folding time of around 4.3 $\mu$s. The ability
of the Markov model to reasonably reproduce the folding time of the original
trajectory is remarkable.  Specifically, it indicates that the kinetic lumping
via network clusterization correctly partitioned the whole
free-energy landscape. When this would not be the case \cite{rao2010}, a much faster kinetics
usually appears \cite{krivov2004, berezovska2012}.

\begin{figure}
\includegraphics[width=80mm,angle=0] {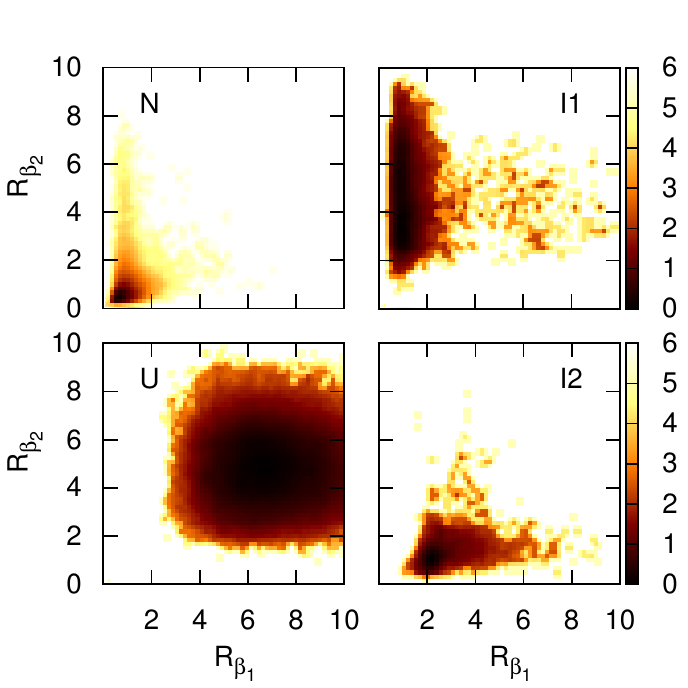}
\caption{Projected free-energy surface of Fip35 for the native, unfolded and the two
intermediate states as found by the local fluctuation analysis. Free energies are expressed in units of $kT$.}
  \label{fig:2Dstates}
\end{figure}

\subsection{Structural analysis of Fip35 folding intermediates}

In Fig.~\ref{fig:2Dstates} a free-energy projection
of the native, unfolded, I1 and I2 states onto the $R_{\beta_1}$ and $R_{\beta_
2}$ coordinates is shown.  The four states occupy well defined regions of the
map. However, the distributions of intermediate states are rather broad producing large overlaps with 
both the native and unfolded states (compare also with Fig.~\ref{fig:projections}
and check references \cite{rao2004,krivov2004}).  Besides this, the two intermediate states
have a good degree of nativeness: for the case of I1 (I2) the value of
$R_{\beta_1}$ ($R_{\beta_2}$) was most of the time below $2$ \AA.

\begin{figure*}
  \includegraphics[width=140mm]{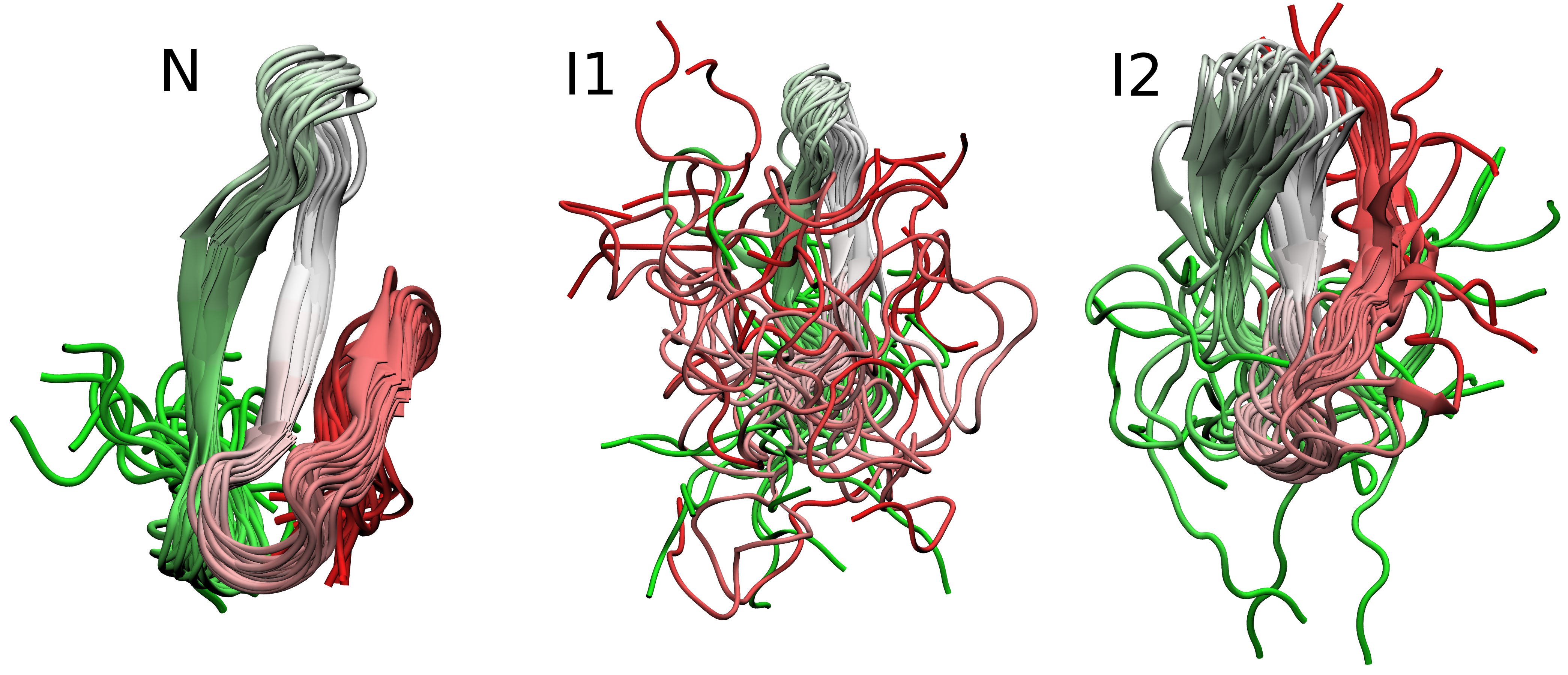}
  \caption{Structural superpositions of  the native and the two
  intermediate states as found by the local fluctuation analysis.  Each  panel
contains 25 randomly chosen frames.  Structures were superimposed on residues
$7-23$ and $18-29$ for I1 and I2, respectively.}
  \label{fig:structures}
\end{figure*}

From a structural point of view, the two intermediates are characterized by
well-defined conformations.  Structural superpositions of the native and
intermediate states are shown in Fig.~\ref{fig:structures} (while the unfolded
state looks like a disordered bundle of structures).  The three states present
a reasonable amount of structural homogeneity. For the case of I1, the first
hairpin is in its native conformation as well as the turn connecting the two
hairpins.  On the other hand the second hairpin is unstructured, interacting
with the rest of the protein in several non-specific ways. In the I2
intermediate, an inverted situation was found where the second hairpin is
native. In this case however, the first hairpin is prone to fold into a
specific configuration instead of being unstructured. This conformation
resembles the native structure but the formation of the first turn, and
consequently the entire hairpin, is shifted by one residue (out-of-register
conformations are typical in $\beta$-hairpins \cite{rao2004}).  

It is striking
to see  the different amount of disorder  in I1 and I2
(Fig.~\ref{fig:structures}), suggesting a qualitative reason for the different
statistical relevance of the two folding pathways. In fact, folding through I1
involves the formation of new specific contacts in the $\beta_2$ region while
folding through I2 first requires the disrupture of a number of non-native
contacts in the $\beta_1$ region due to the out-of-register conformation. As
such, I2 might even be considered per se a misfolded structure.

\section{Discussion}

Complex network analysis of Fip35 RMSD local fluctuations provided evidence for
three main observations:  (i) beyond the native and unfolded states, hidden
states were detected; (ii) among those states, two on-pathway intermediates for
folding were found; (iii) the different amount of structural disorder in the
two intermediates suggest a reason for the prelevance of one pathway with
respect to the other.

Previous calculations based on Markov-state-models found multiple
pathways and a heterogeneous molecular mechanism. In contrast to us, structural
clustering  \cite{pande2011} or likelihood methods in conjunction with contact
maps \cite{hua2012} were used to build the Markov models. Although it is  difficult to
compare these approaches in a quantitative way, their predictions are in qualitative agreement with
our results.  Interestingly, the use of an alternative simulation protocol to
probe slow conformational transitions confirmed the presence of two main
folding pathways \cite{faccioli2012}.   Given the triple stranded native topology of a WW domain, the
presence of these two pathways is not new. The same folding
routes were already observed in the past for a 20 residues triple stranded
$\beta$-sheet peptide in implicit solvent \cite{ferrara2000,rao2004}.  

\begin{figure}
\includegraphics[width=80mm,angle=-90] {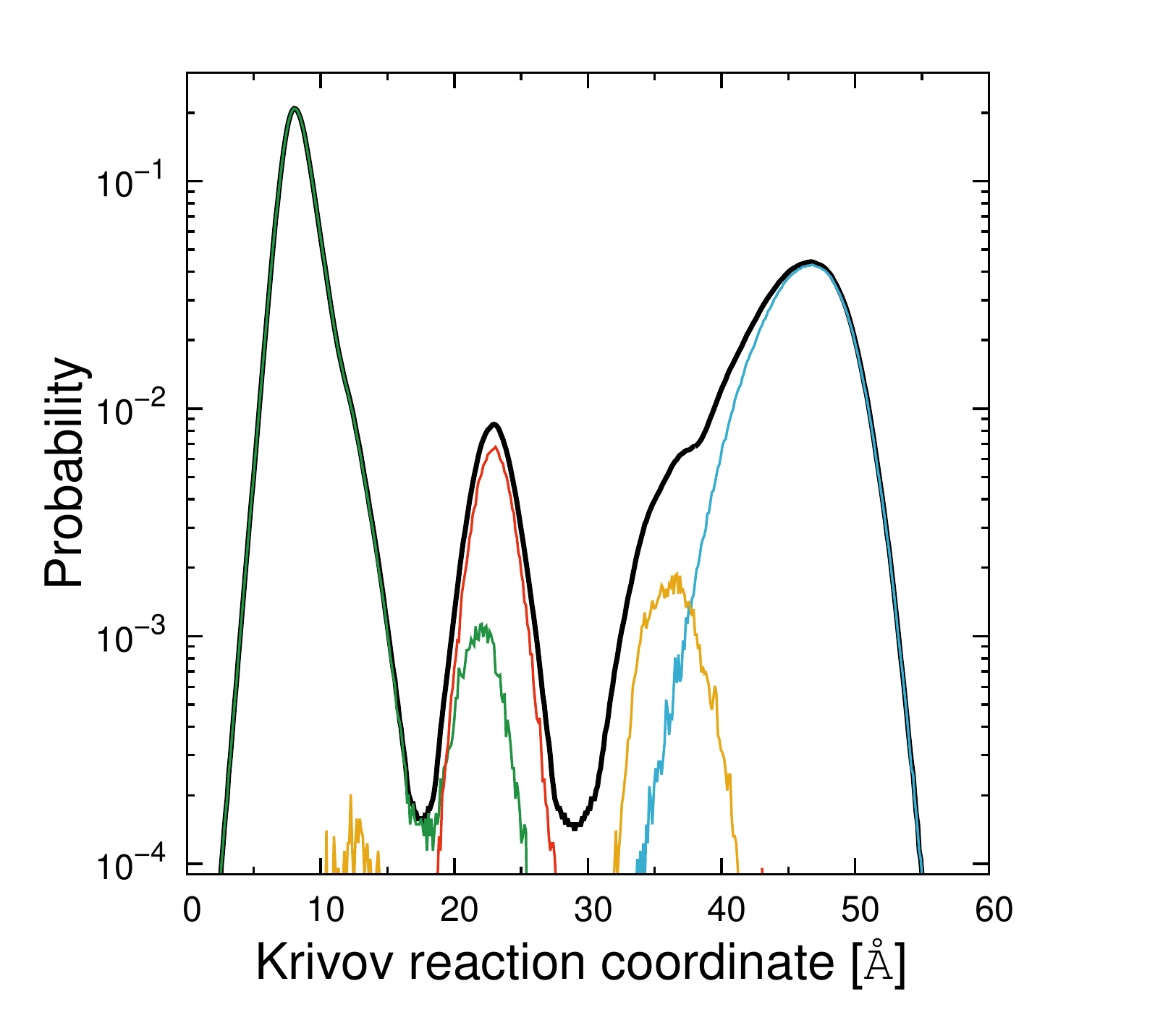}
\caption{Projection of the detected states on the optimal reaction coordinate
developed by Krivov \cite{krivov2011}. Native, unfolded, I1 and I2 states are
shown in green, blue, red and yellow, respectively.}
  \label{fig:krivov}
\end{figure}

So far, one of the most robust interpretations of Fip35 folding 
was provided by the analysis of Krivov \cite{krivov2011}.  Our
findings are in excellent agreement with that study.  This is quantitatively
shown in Fig.\ref{fig:krivov} where we projected our states to the optimized
coordinate developed in that work.  This comparison reveals that the two
approaches provide very similar results.  In Krivov's profile the native, I1
and unfolded states were identified as peaks of the probability distribution
\cite{krivov2011}.  Strikingly, the distributions arising from our detected
states overlap very well with these peaks. For the native state, only a very
small fraction of $0.7 \%$ was found in the wrong part of the profile (green
peak between the value 18 and 26 of the coordinate) while for I1 there is a perfect agreement. Moreover, the
second intermediate I2, which originally could not be directly detected from
the profile, was found in the same position as predicted in
Ref.~\cite{krivov2011} (I2 was hidden because parallel pathways cannot be
simultaneously displayed in this representation).

Overall, the two approaches provided the same mechanistic understanding. This
is encouraging given the strong diversity of the two methods. In fact, Krivov's
reaction coordinate was obtained via an optimization procedure starting from an
educated guess, i.e. a linear combination of conventional (non-optimized)
coordinates. The procedure makes a parameter space search minimizing the flux on
top of the barriers of the initial free-energy projection. Better results were
obtained when using several (i.e.  thousands) of coordinates, e.g.  all
pairwise inter-atomic distances in a protein \cite{krivov2011,steiner2012}.
Unfortunately, this makes the optimization procedure highly non-trivial
\cite{krivov2011numerical}. The strength of this method is to provide
kinetically meaningful free-energy profiles with diffusive dynamics. The
downsides are the intrinsic limitations of 1D profiles to describe parallel
pathways and the non-trivial optimization procedure. The complex network analysis
presented here does not run any optimization algorithm but attempts to detect hidden
states from the time series of a generic coordinate, e.g. the RMSD.  This makes our
method much faster. The downsides are again the need of an educated guess
for the selection of the coordinate to use as well as potentially larger errors
in the resulting kinetic models. 

In conclusion, most of the results presented so far on Fip35 folding point to the same direction. We
believe that the original interpretation provided by Shaw and co-workers of
downhill folding is due to an inappropriate application of the reaction
coordinate optimization they used.  Being based on commitor
probabilities, that approach was designed and tested for two-state processes
only \cite{best2005}.  Given the implicit assumption of a two-state scheme,
application of this protocol to multi-state systems  leads to inaccurate
results.

\section*{Acknowledgments} We are grateful to D.E. Shaw Research for making
their Fip35 data publicly available.  
This work was supported by the Excellence Initiative of the German Federal and
State Governments.

%

\end{document}